\documentclass[conference]{IEEEtran}

\ifCLASSOPTIONcompsoc
  \usepackage[nocompress]{cite}
\else
  \usepackage{cite}
\fi
\usepackage[T1]{fontenc}
\usepackage{tablefootnote}
\usepackage[utf8]{inputenc}
\usepackage{mathrsfs}
\usepackage{stmaryrd} 
\usepackage{tabularx,lipsum,environ,amsmath,amssymb}
\usepackage[boxruled,linesnumbered]{algorithm2e}

\makeatother
\usepackage{cleveref}
\crefname{Lemma}{Lemma}{Lemmas}
\usepackage{booktabs}
\usepackage{algorithm2e}
\usepackage[table,xcdraw]{xcolor}
\usepackage{tabularx,lipsum,environ,amsmath,amssymb}
\usepackage{caption}
\usepackage{algorithm2e}
\usepackage{graphicx}            
\usepackage{amsmath}             
\usepackage{amssymb}
\usepackage{array}
\usepackage{layout}
\usepackage[margin=1.04in]{geometry}
\usepackage{amsmath,amssymb}
\usepackage{gensymb}

\makeatletter
\newsavebox\myboxA
\newsavebox\myboxB
\newlength\mylenA

\newcommand*\xoverline[2][0.75]{%
	\sbox{\myboxA}{$\m@th#2$}%
	\setbox\myboxB\null
	\ht\myboxB=\ht\myboxA%
	\dp\myboxB=\dp\myboxA%
	\wd\myboxB=#1\wd\myboxA
	\sbox\myboxB{$\m@th\overline{\copy\myboxB}$}
	\setlength\mylenA{\the\wd\myboxA}
	\addtolength\mylenA{-\the\wd\myboxB}%
	\ifdim\wd\myboxB<\wd\myboxA%
	\rlap{\hskip 0.5\mylenA\usebox\myboxB}{\usebox\myboxA}%
	\else
	\hskip -0.5\mylenA\rlap{\usebox\myboxA}{\hskip 0.5\mylenA\usebox\myboxB}%
	\fi}
\makeatother

\usepackage{etoolbox}

\makeatletter
\patchcmd{\@maketitle}{\raggedright}{\centering}{}{}
\patchcmd{\@maketitle}{\raggedright}{\centering}{}{}
\makeatother

\newcommand{\twodots}{\mathinner {\ldotp \ldotp}}

\newfont{\bbb}{msbm10 scaled 700}

\newfont{\bb}{msbm10 scaled 1100}
\newcommand{\CC}{\mbox{\bb C}}

\newcommand{\av}{{\bf a}}
\newcommand{\bv}{{\bf b}}

\newcommand{\uv}{{\bf u}}

\newcommand{\vv}{{\bf v}}
\newcommand{\xv}{{\bf x}}

\newcommand{\Dm}{{\bf D}}

\newcommand{\Id}{{\bf I}}

\newcommand{\Sm}{{\bf S}}
\newcommand{\Tm}{{\bf T}}
\newcommand{\Um}{{\bf U}}

\newcommand{\Vm}{{\bf V}}

\newcommand{\Ac}{{\cal A}}

\newcommand{\Tc}{{\cal T}}

\newcommand{\dsf}{{\sf d}}

\newcommand{\diag}{{\hbox{diag}}}

\begin{document}
\captionsetup[figure]{labelfont={},labelformat={default},labelsep=period,name={Fig.}}
\title{RIS-Based Steerable Beamforming Antenna with Near-Field Eigenmode Feeder}

\author{\IEEEauthorblockN{Krishan K. Tiwari, Giuseppe Caire}
	\IEEEauthorblockA{Technische Universit\"at Berlin, 10587 Berlin, Germany}
		Email addresses: {lastname}@tu-berlin.de
}
\maketitle

\begin{abstract}
We present a novel, power and hardware efficient, 
antenna system leveraging the eigenmodes of the over-the-air propagation matrix from an active multi-antenna feeder (AMAF) to a large reflective intelligent surface (RIS), both configured as standard linear arrays and placed in the near field of each other. We demonstrate the flexibility of the proposed architecture by showing that it is capable of generating radiation patterns for multiple applications such as very narrow beams with low side lobes for space-division multiple access communications,  wide-angle beams for short range automotive sensing and sectorial beaconing, and monopulse patterns for radar angular tracking. A key parameter in our design is the AMAF-RIS distance which must be optimized and it is generally much less than the Rayleigh distance. For a given AMAF aperture, the optimal AMAF-RIS distance increases as a function of the RIS size. The AMAF-RIS loss is compensated almost exactly by the larger aperture gain of the RIS leading to almost constant RIS gain with increasing RIS sizes. This allows to choose different beam angular selectivities with the same center beam gain. Active RF amplification is done at the AMAF only, thus resulting in a much higher power efficiency and much lower hardware complexity than conventional phased arrays with same beamforming performance.
\end{abstract}


\IEEEpeerreviewmaketitle

\section{Introduction}
\label{sec:intro}

6G research is pushing the radio spectrum frontier to sub-THz bands \cite{ted_thz} as 5G mmWave systems are already being deployed. Multiuser MIMO technology has enabled large spectral efficiencies in conventional cellular frequency bands (the so-called FR1 tier of sub-6GHz frequencies) via multiuser precoding implemented by baseband digital processing. 
However, this approach does not scale well to very high carrier frequencies and large signal bandwidths due to the very large power consumption of per-antenna A/D conversion. 
On the other hand, mmWave/sub-THz channels are formed by a small number of paths 
with the line of sight (LoS) typically being the strongest. 
Therefore, hybrid digital analog architectures are widely advocated in these bands. 
For example, with a simple ``one stream per subarray'' architecture \cite{commit_bf1}, the number of data streams corresponds to the number of RF chains (A/D conversion), while each data stream is transmitted via beamforming (BF) in the direction of the LoS of the corresponding user. In this case, BF is implemented directly in the RF domain, so that the number of RF chains can be kept quite low, while the size of the BF array can be very large.

\begin{figure}[ht]
\centerline{\includegraphics[width=8.25cm]{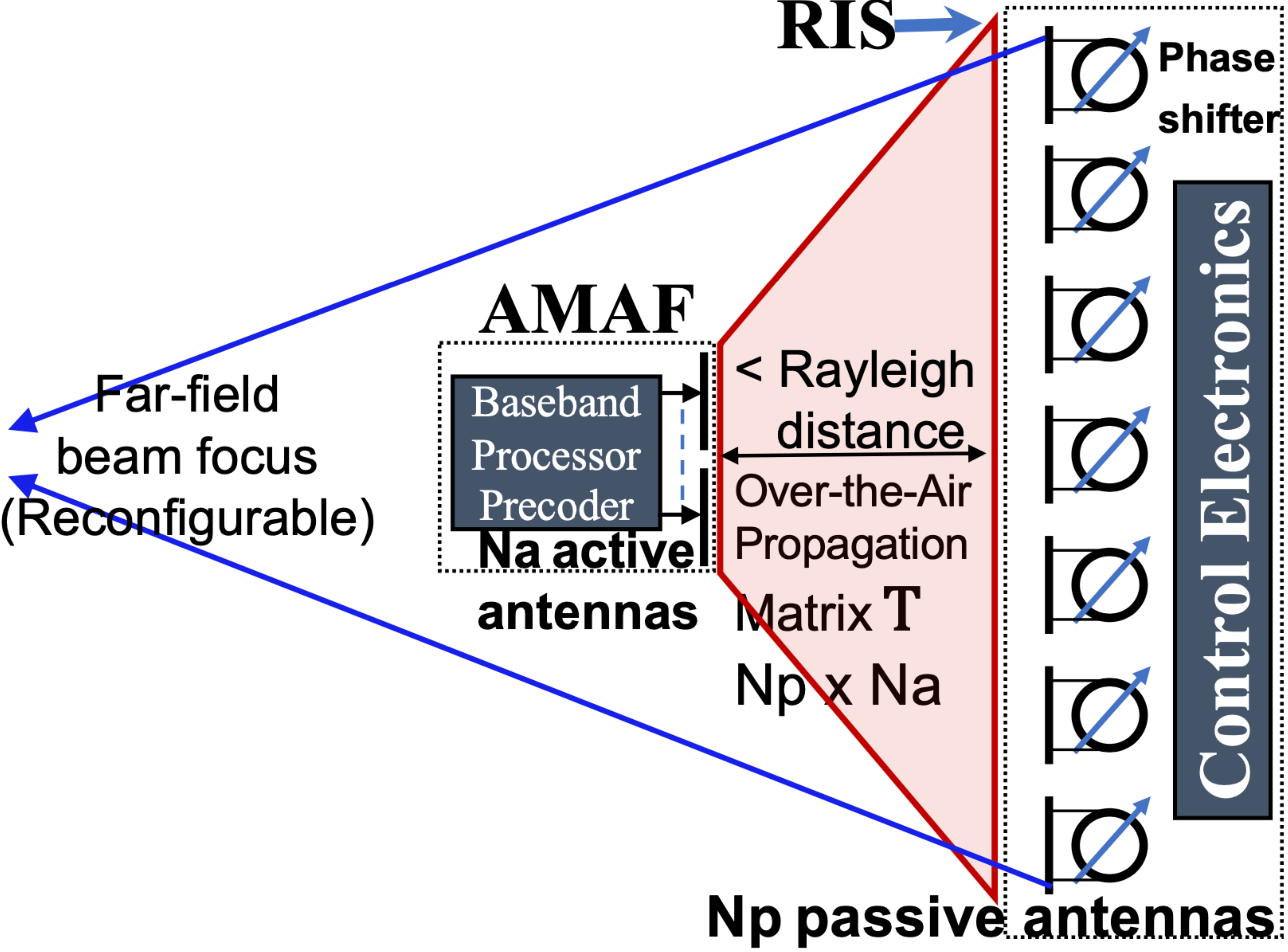}}
	\caption{RIS fed by an AMAF placed in its near field.}
	\label{fig:sysmo}
\end{figure}

The key element of the one-stream per subarray approach is the beam-steering (sub)array. 
In this paper, we focus on an alternative implementation of such array. In the proposed architecture (see Fig.~\ref{fig:sysmo}), a very large, passive Reflective Intelligent Surface (RIS), formed by passive elements with phase-only control and no RF amplification, is fed by an Active Multiantenna Feeder (AMAF) placed in its near field. This is a different use of RIS with respect to the broader literature that considers RIS 
is in the far-field of both the transmitter and the receiver. Our architecture is in line with \cite[Figure 1 (e)]{smt} and with \cite{buzzi}. This can also be seen as a reconfigurable reflectarray \cite{reflectarray} where the horn feeder is replaced by the AMAF. 
The main difference between our present work and \cite{smt,buzzi} is that these works consider a rich scattering channel (which must be estimated) between the RIS elements and the users and some form of digital baseband processing. Therefore, they are more suitable to the FR1 frequency tier. In contrast, here we are interested in the BF capability of the AMAF-RIS structure and consider its radiation pattern seen at a user antenna located in LoS and in the far field with respect to the RIS, which is relevant for mmWave and sub-THz systems. Furthermore, we provide an accurate analysis, albeit under certain simplifying assumptions, of the power efficiency of the proposed structure with respect to a conventional BF array generating the same BF pattern and received power at the target user. Note also that the array fed near field RIS hardware are already being developed and tested, e.g., \cite{Webinar}. 

\textbf{Mathematical notations:} $\Id_\text{N} \in \CC^{N \times N}$ is the identity matrix, $|\text{x}|$ is the magnitude of the scalar $\text{x}$, $|\xv|$ is a vector which contains the magnitudes of the elements of the vector $\xv$, and $[\cdot]^*$ indicates Hermitian transpose. All distances in this paper are normalized to the half wavelength.


\section{System Model}
\label{sec:sys_mo}

Although the typical channel bandwidth for high carrier frequencies may be very large, 
most wireless communication systems are well within the ``narrowband'' condition. 
For example, at a carrier frequency of 150 GHz, even a channel bandwidth of 10 GHz is less than 10\% of the carrier.
Under such assumption, the AMAF to RIS propagation matrix $\Tm \in \CC^{N_p \times N_a}$ ($N_p$ and $N_a$ are sizes of the RIS and the AMAF standard linear arrays (SLAs) \cite[p. 51]{trees}) is given by
\begin{equation}
\label{eq:T}
\centering
T_{n,m} = \frac{\sqrt{E_A(\theta_{n,m})E_R(\phi_{n,m}})~ \text{exp}\big(j\pi r_{n,m}\big)}{\big(2\pi r_{n,m}\big)},
\end{equation}
where $E_A, ~E_R, ~\theta_{n,m},~ \phi_{n,m}, ~ \text{and}~r_{n,m}$ are AMAF element power radiation pattern, RIS element power radiation pattern, angle of departure from an AMAF element $m$ to a RIS element $n$ measured from the AMAF element boresight, angle of arrival from the AMAF element $m$ to the RIS element $n$ measured from the RIS element boresight, and distance (normalized by the half wavelength) between the AMAF element $m$ and the RIS element $n$, respectively. Note that (\ref{eq:T}) is the Friis transmission equation \cite[eq. (2-119)]{Balanis_antenna_theo} in the magnitude form along with the incorporation of the distance dependent phase term.
Note also that, while the AMAF is in the near field of the RIS, the individual antenna elements of the AMAF and the RIS are in far fields of each other.

In this work we assume patch antenna elements (both at the RIS and the AMAF) with axisymmetric power radiation patterns \cite[eq. (2-31)]{Balanis_antenna_theo} given by 
\begin{equation}
\label{eq:patch}
\centering
E_R({\theta})=E_A({\theta})=E({\theta}) = 4~\text{cos}^2(\theta),
\end{equation}
where $\theta$ is the angle from the element boresight. Note that $E({\theta})$ has the half power beam width (HPBW) of 90$\degree$ and the gain of 6 dBi which typical of patch antennas widely used for RIS hardware developments.

The reconfigurable phase shifts at the passive RIS are modelled by a diagonal matrix $\Dm \in \CC^{N_p \times N_p}$ given as $\Dm=\text{diag}(e^{j\phi_1}, \twodots, e^{j\phi_{N_p}})$. Note that if a given beam shape is synthesized with a given RIS configuration $\Dm$, then it can easily be steered to a desired direction because phases are additive. If $\Dm'=\text{diag}(e^{j\phi_1'}, \twodots, e^{j\phi_{N_p}'})$ is the phased-array phase gradient required for steering an arbitrary beam to a given direction, then the RIS configuration for the steered beam shape is given by $\Dm'\Dm=\text{diag}(e^{j(\phi_1'+\phi_1)}, \twodots, e^{j(\phi_{N_p}'+\phi_{N_p})})$.

We assume the user equipment (UE) to be in the far field of the RIS and thereby the RIS far field power radiation pattern (based on the planar wave modelling) is given by 
\begin{equation}
\label{eq:patrn}
\centering
P(\theta)= \left|\av^*(\theta)\Dm\Tm\bv\right|^2~ E_R(\theta),
\end{equation} 
where $\bv \in \CC^{N_a \times 1}$ is an AMAF precoder and the vector $\av(\theta)$ is the RIS array steering vector corresponding to the angle $\theta$ from the RIS broadside and is given by 
\begin{equation}
\label{eq:ARV}
\centering
\av(\theta)=\left[1,e^{j \pi \text{sin}(\theta)},\twodots,e^{j \pi (N_p-1)\text{sin}(\theta)}\right]^*.
\end{equation}

\section{Eigenmode-based Designs}  
\label{sec:eig_des}

We consider SVD of $\Tm=\Um\Sm\Vm^*$ where $\Um \in \CC^{N_p \times N_p}$ and $\Vm \in \CC^{N_a \times N_a}$ are unitary matrices and $\Sm \in \CC^{N_p \times N_a}$ is a diagonal matrix containing ordered singular values $\sigma_1, \sigma_2, \twodots, \sigma_{N_a}$. $\uv_i$ and $\vv_j$ are the $i^{\text{th}}$ and the $j^{\text{th}}$ column vectors of $\Um$ and $\Vm$, respectively.

We present eigenmode-based designs where the AMAF precoder $\bv$ is given as 
\begin{equation}
\centering
\bv= \sum_{i=1}^{N_a} \beta_i \vv_i,  \label{suca}
\end{equation} 
where $\beta_i \in \mathbb{R}$ and $\mathbb{R}$ is the set of real numbers. An eigenmode-based design leads to a RIS beamforming vector given as
\begin{equation}
\label{eq:OTA_bf}
\centering
\Dm \sum_{i=1}^{N_a} \sigma_i \beta_i \uv_i \in \Ac,
\end{equation} 
where $\Ac$ is the set of all eigenmode-based designs. Note that since the columns of $\Vm$ forms a unitary basis for $\CC^{N_a}$, there is no loss of generality to represent the AMAF BF vector
$\bv$ in the form (\ref{suca}). Also, the corresponding excitation vector  
$\sum_{i=1}^{N_a} \sigma_i \beta_i \uv_i$ over the RIS elements is any vector in the linear subspace $\Tc = \text{Span}(\Tm)=\text{Span}(\uv_1, \twodots,\uv_{N_a})$. However, 
$\Ac$ is not a linear subspace, since it contains all element-wise phase rotated vectors in $\Ac$, by the action of the matrix $\Dm$. Mathematically, $\Ac$ is the orbit of the 
subspace $\Tc$ under the multiplicative (infinite) unitary group of diagonal matrices with unit-modulus diagonal elements. A non-trivial problem consists of the best approximation of some desired target BF vector by elements of $\Ac$. Instead of addressing this problem, which seems to be quite difficult, we consider here some special choices of elements in $\Ac$ with attractive properties from a communication viewpoint. 

\subsection{High Gain Very Low Sidelobe Patterns}  
\label{subsec:hg}

\begin{figure}[ht]
\hspace{-10pt}
	\includegraphics[width=8.15cm]{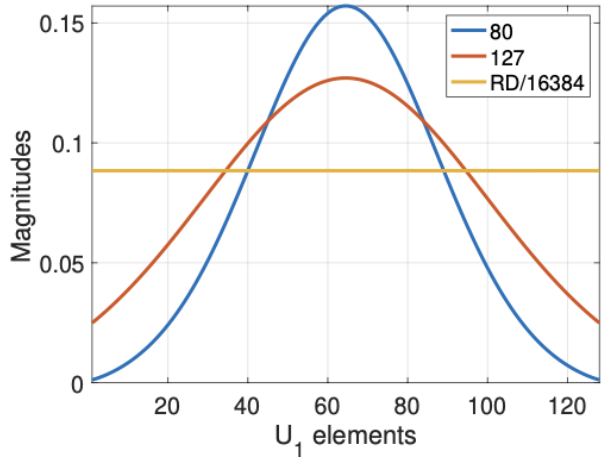}
	\caption{Natural tapering in $|\uv_1|$ at $\dsf = 80, 127,~ \text{and}~ 16384$. $N_p=128,~ N_a=4, ~\text{Rayleigh distance (RD)}=N_p^2=16384$.}
	\label{fig:u1amp}
\end{figure}

For maximum power transfer from the AMAF to the RIS for a given AMAF-RIS distance $\dsf$, the AMAF must be precoded along the principal eigen mode (PEM) of $\Tm$, i.e., $\bv=\vv_1$. 
In order to obtain maximum gain in the RIS broadside direction (notice that the max-gain direction can be shifted by standard beam steering as said before), all elements of the resulting excitation vector at the RIS elements, given by $\Tm \vv_1 = \sigma_1 \Dm \uv_1$ 
must be real and positive. Hence, we choose $\Dm$ such that $\Dm \uv_1$ has real positive elements,\footnote{For simplicity, we are considering continuous phase-shifters at the RIS in this work. An investigation of the impact of discrete phase shift values is a future action item.} and denote the resulting vector as 
$\Dm \uv_1 = |\uv_1|$. 
As seen in Fig. \ref{fig:u1amp}, for $\dsf$ much less than the Rayleigh distance (RD), a natural tapering is seen in $|\uv_1|$. This tapering is due to: (i) larger distance dependent propagation losses towards RIS ends than at its center, (ii) smaller element pattern gains for the larger angles at array ends than at the center array elements which see each other at their boresight.

For $\dsf < 80$, $\sigma_1|\uv_1|$ exhibits multiple peaks and for $\dsf > 80$ the AMAF-RIS propagation coefficient $\sigma_1^2$ decreases quadratically with $\dsf$ in free space, therefore $\dsf=80$ is an optimum value.

\begin{figure}[ht]
\centering
\hspace{-10pt}
	\includegraphics[width=8.15cm]{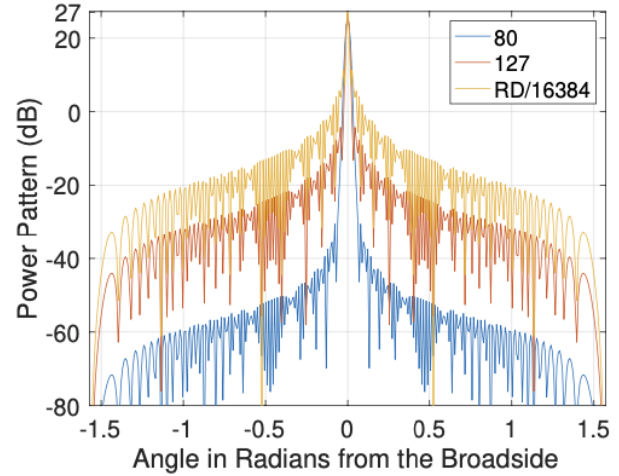}
	\caption{Far field radiation patterns $\big|\av^*(\theta)|\uv_1|\big|^2 E(\theta)$ at different AMAF-RIS distances $\dsf = 80, 127,~ \text{and}~ 16384$. $N_p=128,~ N_a=4, ~\text{Rayleigh distance (RD)}=16384$.}
	\label{fig:u1pat}
\end{figure}

\begin{figure}[ht]
\centering
	\includegraphics[width=8.15cm]{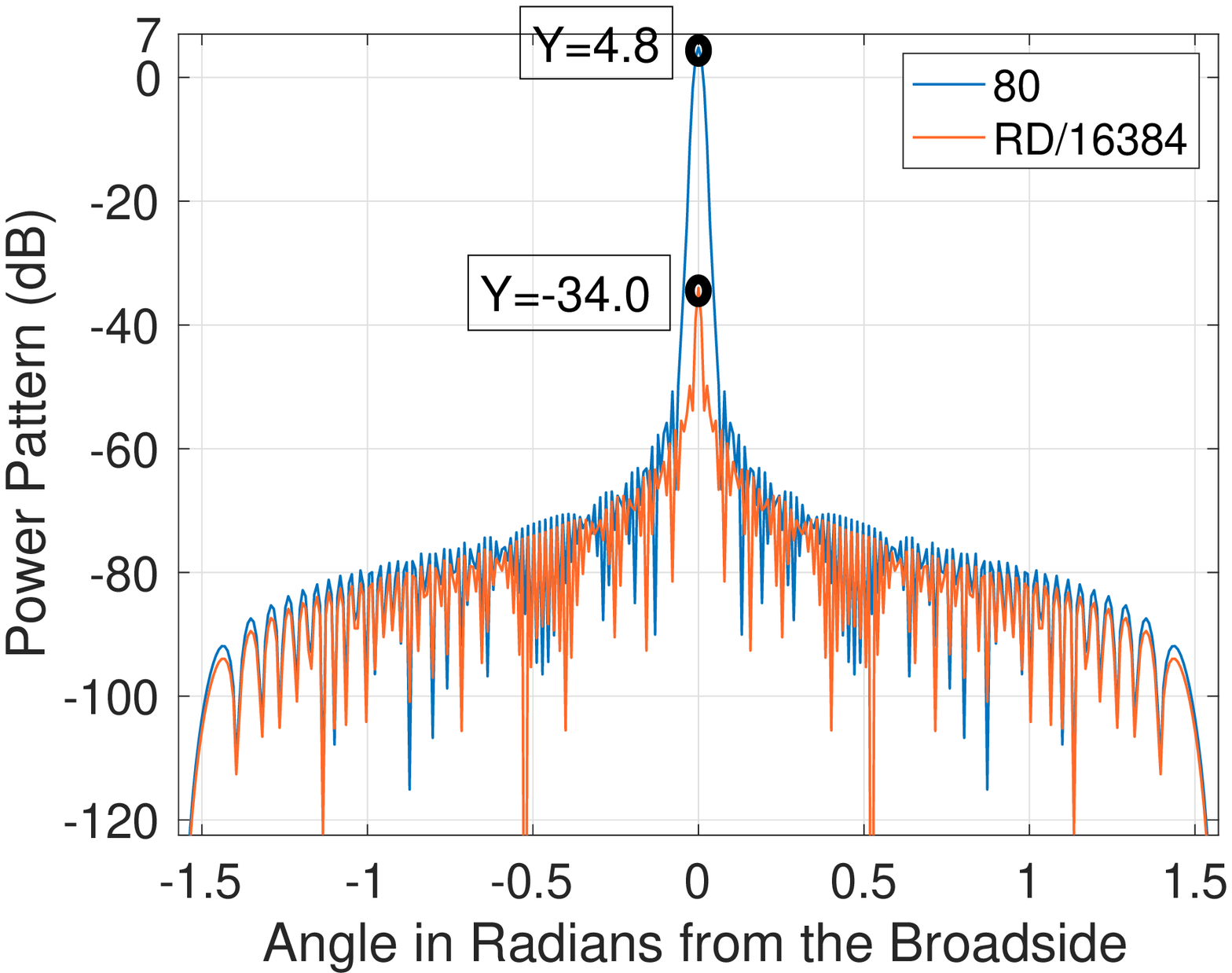}
	\caption{Merit of design 1 in terms of the far field patterns $\big|\av^*(\theta)\Dm_1 \Tm \vv_1\big|^2 E(\theta)$.}
	\label{fig:PEM_Gain}
\end{figure}

Thus, the first eigenmode based design (design 1) proposed here is $\dsf=80$ (for $N_a=4$ and $N_p=128$), $\bv_1 = \vv_1$, and $\Dm_1 = \diag(u_{11}^*/|u_{11}|, \ldots, u_{1N_p}^*/|u_{1N_p}|)$. Ignoring the factor $\sigma_1^2$, we see in Fig. \ref{fig:u1pat}, that design 1 achieves sidelobe rejection better than 55 dB with only a small reduction (2 dB) in the main lobe peak with reference to the Rayleigh distance (RD); this 2dB reduction in the main lobe peak is more than compensated by an increase of 41 dB in $\sigma_1^2$ at $\dsf=80$ as compared to the RD. Overall, as seen in Fig.~\ref{fig:PEM_Gain}, design 1 yields 39 dB stronger signal at the far field UE with almost the same sidelobe levels as for the case where the AMAF is placed at the RD of the RIS.

\subsection{Wide Angle Flat Top Patterns}  
\label{subsec:fl}

The eigenmodes of $\Tm$ demonstrate very large magnitude taperings across the large RIS array as we have already seen for the principal eigenmode of $\Tm$ in sub-section \ref{subsec:hg}. These natural magnitude taperings in the eigenmodes of $\Tm$ can be leveraged for syntheses of different beam shapes by various linear combinations of different eigenmodes.

As a proof of concept, we now present design 2 with $\dsf=80$ (for $N_a=4$ and $N_p=128$), $\bv_2=2~\vv_1/\sigma_1 + \vv_3/\sigma_3$, and $\Dm_{2}=\Id_{N_p}$. Note that while design 1 required phase compensations at the RIS, $\Dm_{2}=\Id_{N_p}$ meaning this design is free-of-cost in the sense that it does not require any additional processing or overheads other than just precoding the AMAF with $\bv_2$ which is very simple.

\begin{figure}[ht]
\centering
	\includegraphics[width=8.15cm]{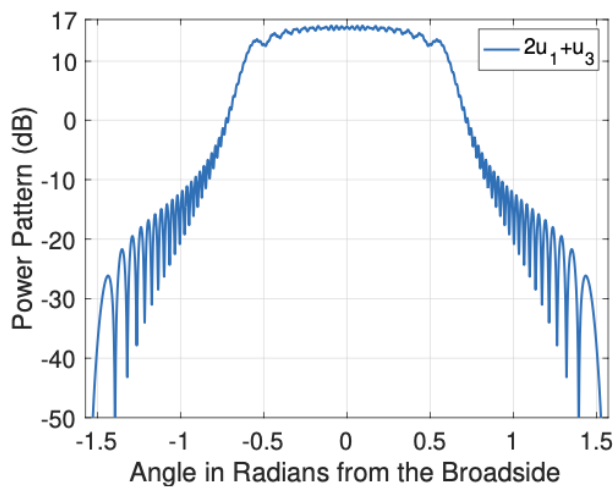}
	\caption{Flat top beam pattern synthesis using design 2.}
	\label{fig:flat_top}
\end{figure}

As seen in Fig.~\ref{fig:flat_top}, design 2 yields a beam pattern with a quite flat gain of $>15~ \text{dBi}$ in the sector of $+/-15\degree$ from the RIS broadside and $>13.7~ \text{dBi}$ in the sector of $+/-28\degree$ after which it falls off rapidly with the angle. This design 2 can be used for a wide-angle illumination beam for short-range automotive radar or for a wide-angle beacon signal in order to identify the cell and allows the UE to acquire the random access channel slot.

\subsection{Monopulse Patterns}  
\label{subsec:MONOPULSE}
In IEEE Standard 145-2013 \cite[p. 22]{145-2013}, monopulse is defined as ``simultaneous lobing whereby direction-finding information is obtainable from a single pulse.'' The AMAF needs to have at least two RF chains for monopulse angular tracking - one for the difference pattern and for the sum pattern \cite{monopulse_book}.

The proposed design 3 consists of the choices $\Dm_3=\Id_{N_p}$, $\bv_{3a}=\vv_1$ for the sum and $\bv_{3b}=\sigma_1\vv_2/\sigma_2$ for the difference. $N_p=128$, $N_a=4$. Note that the scaling factor of $\sigma_1/\sigma_2$ has been used in $\bv_{3b}$ in order to equalize for the variations due to the difference between $\sigma_1$ and $\sigma_2$ so that the monopulse sum and difference patterns have the same scale. Further, if the AMAF implements fully digital baseband beamforming with $N_a$ RF chains, then this can be easily implemented in the baseband without the need of any extra hardware. If the AMAF implements analog beamforming with only one RF chain, then one AMAF can realize either $\bv_{3a}$ or $\bv_{3b}$. In that case, two close-by AMAFs can be used to draw the monopulse sum signal and the monopulse difference signal corresponding to $\bv_{3a}$ and $\bv_{3b}$, respectively. Because the sum signal is used only for the normalization of the difference signal to make the monopulse error signal independent of the range, the normalization can very well be done by the $\bv_{3a}$ output signal from an additional close-by AMAF.

\begin{figure}[ht]
\hspace{-5pt}
	\includegraphics[width=8.15cm]{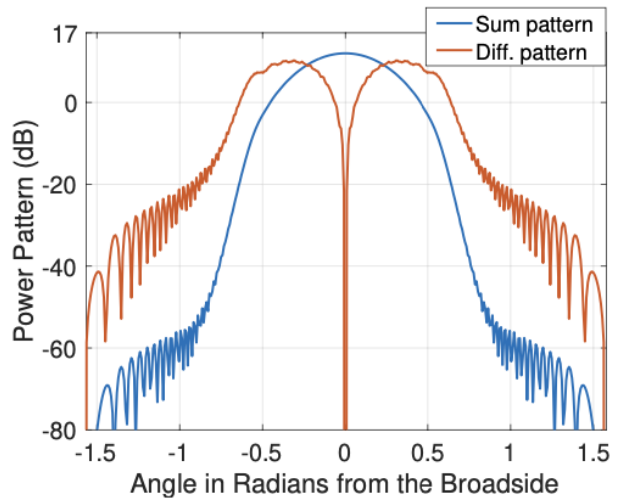}
	\caption{Monopulse pattern from design 3 at $\dsf=80$.}
	\label{fig:monopulse_v1v2}
\end{figure}

In Fig. \ref{fig:monopulse_v1v2}, we see that the null depth is more than 80 dB which enables a very good D/S monopulse error curve with a good slope for an accurate angular tracking. Conventionally, a monopulse system needs a dedicated complex monopulse comparator hardware. In contrast, our architecture of Fig.~\ref{fig:sysmo} can cater to the same functionality by using the principal and the second eigenmodes of $\Tm$. For an implementation with even only two RF chains, no extra hardware will be needed.

\section{RIS Gain vs RIS Size, Element Gain, $\&$ AMAF Size}
\label{sec:GvsS}

\begin{figure}[ht]
\hspace{-8pt}
	\includegraphics[width=8.15cm]{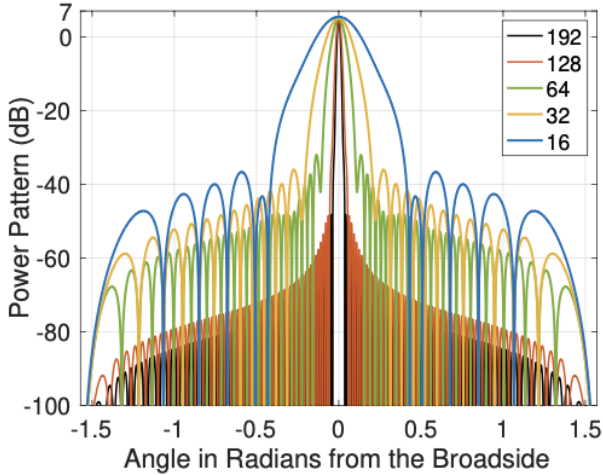}
	\caption{Different beamdwidths with the same gain from different RIS sizes.}
	\label{fig:sum}
\end{figure}

In Table \ref{tab:TRG}, we report results for design 1 (principal eigemode BF) 
for different RIS size $N_p$ and fixed AMAF size $N_a=4$. We see that the optimal AMAF-RIS distance $\dsf$ increases with RIS size $N_p$ such that the AMAF-RIS loss $\sigma_1^2$ is compensated almost exactly by the larger aperture gain of the RIS given as $4\big|\av^*(0)\big|\uv_1\big|\big|^2$. 
The factor of 4 is due to the element gain $E(0)=4$ from (\ref{eq:patch}). It follows that we obtain roughly same total center-beam RIS gain (given by $\Gamma=\sigma_1^2 \big|\av^*(0)\big|\uv_1\big|\big|^2E(0)$) for different beam angular selectivities, by choosing different RIS sizes $N_p$ as seen in Fig. \ref{fig:sum}. 

\begin{table}[ht]
\caption{RIS Gain $\Gamma$ vs RIS sizes $N_p$, with $E(\theta)$}
\label{tab:TRG}
\centering
\setlength{\tabcolsep}{1.5pt} 
\begin{tabular}{c c c c c c c c  }
\hline\hline 
$N_p$  & $\dsf$   & \begin{tabular}[c]{@{}c@{}} $\Gamma$\\dBi \end{tabular} & \begin{tabular}[c]{@{}c@{}}$4\big|\av^*(0)\big|\uv_1\big|\big|^2$\\dB\end{tabular} & \begin{tabular}[c]{@{}c@{}}$\sigma_1^2$ \\ dB\end{tabular} & \begin{tabular}[c]{@{}c@{}}$\sigma_1\big|\uv_1\big|$ \\ Taper (dB)\end{tabular}   & \begin{tabular}[c]{@{}c@{}}SLL\\dB\end{tabular} & \begin{tabular}[c]{@{}c@{}}$\big|\vv_1\big|$ \\ Taper (dB)\end{tabular}  \\
\hline 
16  & 8   & 5.4 & 15.7 & -10.3 & 13.4 & -42.0  & 2.5 \\
32  & 16  & 5.1 & 18.3 & -13.3 & 15.4 & -41.0 & 2.5 \\
64  & 30  & 5.0 & 20.9 & -16.0 & 17.4 & -36.9 & 2.5 \\
128 & 80  & 4.8 & 25.0 & -20.2 & 21.1 & -55.6 & 2.6 \\
192 & 120 & 4.8 & 26.8 & -22.0 & 22.1 & -55.3 & 2.6 \\   \hline 
\end{tabular}
\end{table}

Let us recall that for parabolic reflectors, the reflector gain increases with increasing reflector size. The feed is placed at the focal point of the parabolic reflector. For an increased parabolic reflector size, the distance of the focal point from the center of the parabolic reflector does not change. However, in our case, the optimum AMAF-RIS distance $\dsf$ increases as $N_p$ increases. This is a fundamental difference in the effect of the increased reflector size for a parabolic reflector versus a ``flat reflector.''

Note from Table \ref{tab:TRG} that the natural taper in $\sigma_1|\uv_1|$ generally increases with increasing $N_p$ which causes improved sidelobe rejections for larger RIS sizes. However, the natural taper in $\big|\vv_1\big|$ doesn't change much because the AMAF size remains constant at $N_a=4$. In Table 1, the f/D ratio (focal distance `f' is the distance between the AMAF and the RIS; `D' is the RIS size) is less (0.5 to 0.66) than the typical value of 0.9 to 1 for horn fed reflectarrays. This is because of the principal eigenmode BF which enables the maximum power transfer from the AMAF to the RIS and the discretized feed aperture which is continuous for the horn feed with minimal taper. 

By (\ref{eq:T}) $\Tm$ is multiplied by $E(0)$ because we consider the same antenna element at the AMAF and the RIS. In addition, the increased element gain also manifests in the far field pattern from (\ref{eq:patrn}). This means that an increase of $\gamma$ dB in $E(0)$ should yield an increase of $3\gamma$ dB in the RIS gain $\Gamma$. With this motivation, we re-ran the numerical experiments wherein we regenerated $\Tm$ of (\ref{eq:T}) with the antenna element model $E_h(\theta)=6.3~ \text{cos}^4(\theta)$ having the HPBW of 65.5$\degree$ and the gain of 8 dBi which is close to the model of \cite[Table 7.3-1]{3GPP}. The results have been consolidated in Table \ref{tab:TRG_h}.   

\begin{table}[ht]
\caption{RIS Gain $\Gamma$ with RIS sizes $N_p$, with $E_h(\theta)$}
\label{tab:TRG_h}
\centering
\setlength{\tabcolsep}{3pt} 
\begin{tabular}{c c c c c c c}
\hline\hline 
$N_p$  & $\dsf$   & \begin{tabular}[c]{@{}c@{}} $\Gamma$\\dBi \end{tabular} & \begin{tabular}[c]{@{}c@{}}$6.3\big|\av^*(0)\big|\uv_1\big|\big|^2$\\dB\end{tabular} & \begin{tabular}[c]{@{}c@{}}$\sigma_1^2$ \\ dB\end{tabular} & \begin{tabular}[c]{@{}c@{}}$\sigma_1\big|\uv_1\big|$ \\ Taper (dB)\end{tabular}   & \begin{tabular}[c]{@{}c@{}}SLL\\dB\end{tabular}\\
\hline 
16  & 8   & 10.8 & 17.5 &  -6.6  & 13.8 & -42.8   \\
32  & 16  & 10.4 & 20.0 & -9.6  & 16.0 & -40.9  \\
64  & 30  & 10.5 & 22.3 & -12.3  & 16.8 & -49.1  \\
128 & 80  & 10.2 & 26.6 & -16.6  & 21.6 & -57.2  \\
192 & 120 & 10.3 & 28.4 & -18.4 & 23.1 & -56.8  \\   \hline 
\end{tabular}
\end{table}

As we can see in Table \ref{tab:TRG_h}, for all $N_p$ values, the corresponding optimum AMAF-RIS distances ($\dsf$) remained the same as in Table \ref{tab:TRG}, i.e., the f/D sensitivity on AMAF element gain is much lower than that for horn fed reflectarrays (recall that a higher gain horn feed is typically placed at a larger f/D ratio than a lower gain horn). Further, the RIS gain $\Gamma$ is again almost flat with increasing RIS sizes $N_p$ in Table \ref{tab:TRG_h}. Note that the RIS gain $\Gamma$ has indeed increased by almost 5.5 dB from the corresponding $\Gamma$ values in Table \ref{tab:TRG}. We expected an increase of 6 dB in $\Gamma$, in the previous paragraph, because of the increase of 2 dB in $E(0)$ from $E(\theta)$ of (\ref{eq:patch}) for Table \ref{tab:TRG} to $E_h(\theta)$ for Table \ref{tab:TRG_h}. The anticipated  increase of 6 dB in the RIS gain $\Gamma$ was due to the improvement of 4 dB in $\sigma_1^2$ of $\Tm$ and 2 dB in the RIS element gain post reflection from the RIS. However, let us note that the taper in $\sigma_1\big|\uv_1\big|$ in Table \ref{tab:TRG_h} has typically increased as compared to that in Table \ref{tab:TRG}. Intuitively, this is because the center RIS elements receive more power from the AMAF when the AMAF elements have more directive gain with the narrower radiation pattern $E_h(\theta)$ than when the element pattern $E(\theta)$ was broader in (\ref{eq:patch}). Due to the increased tapers in $\sigma_1\big|\uv_1\big|$ in Table \ref{tab:TRG_h}, the increase of 4 dB in $\big|\av^*(0)\big|\uv_1\big|\big|^2$ in the fifth column has reduced as by about 0.5 dB as compared to Table \ref{tab:TRG}. Therefore, the RIS gain $\Gamma$ increased only by about 5.5 dB instead of the 6 dB we had anticipated. In short, the RIS gain $\Gamma$ is approximately cubic in the element antenna gain $E(0)$, while for classical arrays such as the phased array gain is linear in the element antenna gain $E(0)$. This cubic behavior is seen because we have considered same element factors at the AMAF and the RIS. If the AMAF element gain $E_A(0)$ is different than the RIS element gain $E_R(0)$, then the RIS gain 
\begin{equation}
\label{eq:ele_gains}
\centering
\Gamma \propto E_A(0) E_R(0)^2.
\end{equation}
Note from (\ref{eq:ele_gains}) that higher the directive gain of the AMAF array elements, larger the RIS gain $\Gamma$. The fixed AMAF is expected to point to the fixed RIS, i.e., much steering is not required from the AMAF. Had the AMAF been required to steer its beam to a very wide scan space, then to minimize cusping losses due the inherent angular selectivity of the directive antenna element radiation pattern \cite{VTC2022-Fall}, a broad pattern antenna element would have been desired. However, this is not the case with our fixed AMAF which space feeds the fixed RIS. Therefore, antenna elements with high directive gains are more suitable for the constituting the AMAF for high RIS gains $\Gamma$ by (\ref{eq:ele_gains}). 

On the other hand, for RIS which serve mobile UE, the RIS elements should not have very high directive gains or very narrow beams. This is to enable the RIS to steer its beam in wider scan spaces with minimal cusping losses due to the angular selectivity of the RIS element pattern \cite{VTC2022-Fall}. However, we also see from (\ref{eq:ele_gains}) that the RIS gain $\Gamma \propto E_R(0)^2$. Thus, there is a trade-off in the choice of the RIS element gain $E_R(0)$ such that from the RIS gain $\Gamma$ point of view the element gain $E_R(0)$ should be as high as possible while from the beam steerability point of view broader element patterns enable lower cusping losses due to the angular selectivity of the RIS element factor. Note that the RIS response falls off only linearly with the reduction in the element antenna response at larger offset angles from the RIS broadside. If a RIS is used for a backhaul link with minimal beam steering requirements only to compensate for misalignments due to wind load, etc., then clearly the RIS too should be constituted of high gain antenna elements.

We also investigated the impact of varying the AMAF size $N_a$. As $N_a$ is increased, the natural tapering in $\sigma_1\big|\uv_1\big|$ increases. This is because more feeder elements reinforce their gains in the RIS center which is closer to their boresights than the RIS ends. This leads to a kink in the $\sigma_1\big|\uv_1\big|$ tapering. This disturbs the far-field pattern of the RIS causing beam broadening and spikes in the beam walls. This happens at the optimum $\dsf$ similar to the multiple peak $\sigma_1\big|\uv_1\big|$ tapering we had mentioned in sub-section \ref{subsec:hg} when the AMAF gets too close to the RIS. When $N_a$ is decreased, the $\sigma_1\big|\uv_1\big|$ taper decreases. As a result, the optimum $\dsf$ for $N_p=128$ and $N_a=2$ is decreased to 40. Note that clearly the optimum AMAF-RIS distance $\dsf$ and thereby the f/D ratio is sensitive to $N_a$. Further, due to more pronounced tapers at the $\dsf=40$, the factor $\big|\av^*(0)\big|\uv_1\big|\big|^2$ decreases by 1 dB while the $\sigma_1^2$ increases by 3.2 dB, leading to an increase of 2.2 dB in the RIS gain $\Gamma$. On the other hand, the sidelobe level is poorer and in general the sidelobes get higher by about 20 dB as compared to the best result we had seen earlier with $N_a=4$, $N_p=128$, and $\dsf=80$ in Fig. \ref{fig:u1pat}. Thus, the AMAF size $N_a=4$ not only yields the best RIS far field pattern as in Fig. \ref{fig:u1pat} but also allows four eigenmodes of $\Tm$ enabling multiple functionalities with the designs 2 and 3.

\section{Power Efficiency Analysis}  
\label{sec:eff}
We now present power efficiency comparison of our architecture of Fig.~\ref{fig:sysmo} (arch. 1) against the conventional RF beamforming or the phased array architecture (arch. 2). As an illustrative example, we consider the system requirement specifications of Table \ref{tab:SRS}. The carrier frequency of 100 GHz and the range of 20 m translate to the free space isotropic path loss of $32.5 + 20~ \text{log}_{10} (10^5 ~\text{MHz}) + 20~ \text{log}_{10} (20~ \text{m})=$ 98.5 dB. Recall that the thermal noise power spectral density is -174 dBm/Hz at 25\degree C. With 5 GHz bandwidth and 5 dB receiver (Rx) noise figure, the Rx noise power is $-174~ \text{dBm/Hz}~ + 10~\text{log}_{10} (5*10^9 ~\text{Hz})+5=$ -72 dBm. Let the required Rx signal to noise ratio (SNR) be 3 dB. This means that the Rx signal power should be $-72~ \text{dBm} + 3~ \text{dB}=$ -69 dBm. With the isotropic path loss of 98.5 dB, the Tx RF power should be $P_T=-69~ \text{dBm} + 98.5~ \text{dB} =$ 29.5 dBm.

\begin{table}[h]
\caption{An example system req. specifications}
\label{tab:SRS}
\centering
\setlength{\tabcolsep}{2pt} 
\begin{tabular}{lclc}
\hline\hline 
Specification & Value & Specification & Value \\
\hline 
Carrier freq. (GHz)       & 100 & BW (GHz) & 5   \\
Range (m)                 & 20  & FSPL (dB)  & 98.5 \\
Ther. noise pow. (dBm) & -77 & Rx NF (dB)      & 5 \\
Rx noise pow. (dBm)      & -72  & Req. Rx SNR (dB)      & 3   \\
Rx sigl. pow. (dBm)      & -69 & Tx power (dBm)            & 29.5  \\
Sidelobe level (dB)       & -56  & Array size             & 128\\
\hline 
\end{tabular}
\end{table}

In order to accomplish the same beam selectivity, we keep the transmit SLA size (RIS for arch. 1 of Fig. \ref{fig:sysmo}) to be 128 for both the architectures. Further, we want both the architectures to achieve the same sidelobe level of -56 dB as in Fig. \ref{fig:u1pat}. This, in turn, means the same tapering profile of Fig. \ref{fig:u1amp} ($\dsf=80$) for both the architectures. For arch. 1, the RF power per PA is given by
\begin{eqnarray}
\centering
P_{A1}\!\!\!\!\!&=&\!\!\!\!\!P_T\!-\!10\text{log}_{10}\! \left(\!E_R(0)\sum_{i=1}^{N_p}\big|u_{1i}\big|^2\right)\!+\!10\text{log}_{10}(\sigma_1^2),
\label{eq:amp1a}\nonumber\\
&=&29.5-25.0+20.2=24.7~ \text{dBm},
\label{eq:amp1b}
\end{eqnarray} 
where $10~\text{log}_{10} \big(\sum_{i=1}^{N_p}\big|u_{1i}\big|^2\big)\!\!=\!\!19~ \text{dB}$ for the given taper and $10~\text{log}_{10}(E_R(0))=6~ \text{dBi}$ from (\ref{eq:patch}).

For arch. 2 with the same taper, the RF power per PA is given by
\begin{eqnarray}
\centering
P_{A2}&=&P_T-10~\text{log}_{10} \left(E_R(0) \sum_{i=1}^{N_p}\big|u_{1i}\big|^2\right ),
\label{eq:amp2a}\\
&=&29.5-25.0=4.5~ \text{dBm}.
\label{eq:amp2b}
\end{eqnarray} 

\begin{table}[h]
\caption{State-of-the-art power amplifiers from semiconductor technologies.}
\label{tab:tech}
\centering
\begin{tabular}{cccc}
\hline\hline 
Tech. & Psat (dBm) & PAE (\%) & Gain (dB) \\
\hline 
CMOS (Bulk/SOI\tablefootnote{Silicon on Insulator (SOI)})  & 14         & 10       & 13        \\
SiGe  & 16         & 14       & 14        \\
GaN   & 27         & 14       & 20        \\
GaAs  & 27         & 12.5     & 27        \\
InP   & 20         & 22       & 28        \\
\hline 
\end{tabular}
\end{table}

In Table \ref{tab:tech}, we have listed state-of-the-art power amplifiers (PAs) with the highest RF output power and the power added efficiency (PAE) at 100 GHz extracted from an exhaustive PAs survey \cite{GTU_PA_Survey}. Note that the Gallium Nitride (GaN) technology provides the highest saturated RF output power (Psat) of 27 dBm and the highest PAE of 14\%. Therefore, we take this as the reference for our power efficiency analysis. Due to the 20 dB PA gain, it's practically reasonable to neglect the small signal power consumption before the PA stage for simplicity. Recall that typically the PAs are operated in the linear region with a suitable back-off. The back-off is typically larger for high peak to average power ratio (PAPR) waveforms (e.g., orthogonal frequency division modulation) than for lower PAPR waveforms (e.g., single carrier (SC) modulation). While $P_{A2}$ is easily achieved by all the technologies, $P_{A1}$ is challenging to achieve in the linear region even by the GaN technology in view of a certain back-off requirement (even for SC modulation schemes). 

The PAE of a power amplifier is given as
\begin{equation}
\label{eq:PAE}
\centering
\text{PAE}=\frac{P_{RF}}{P_{DC}} 100~ \%,
\end{equation}
where $P_{RF}$ is the output RF power minus the input RF power and $P_{DC}$ is the total direct current (DC) power consumed. By (\ref{eq:PAE}), the $P_{DC}$ of the GaN PA is 3.54 W for the $P_{RF}=$ 500 mW and PAE = 14 $\%$ from Table \ref{tab:tech}. The conventional RF beamforming architecture with 128 antenna elements needs 128 PAs because typically the PA is the last RF block after the signal splitting network to compensate for the signal splitting and insertion losses. Therefore, the DC power consumed by arch. 1 is $128 \times 3.54 =$ 453.1 W. The RIS-based arch. 2 needs only 4 PAs at the AMAF, thereby it consumes only $4 \times 3.54 =$ 14.2 W of DC power. Clearly, if the $P_{A1}$ specification can be met, then the arch. 1 power consumption is orders of magnitude smaller than the arch. 2 power consumption as seen in Table \ref{tab:comp}. 

\begin{table}[ht]
\caption{Power consumption comparison of RF and RIS-based beamforming architectures.}
\label{tab:comp}
\centering
\setlength{\tabcolsep}{2pt} 
\begin{tabular}{ccc}
\hline\hline 
    & RF beamfmg. & RIS-based beamfmg.\\
\hline 
Power amplifiers (No.) & 128            & 4                     \\
Total DC power (W)   & 453.1            & 14.2       \\  
\hline 
\end{tabular}
\end{table}

Beyond the much higher power efficiency, 
it should also be mentioned that the proposed arch. 1 has a higher hardware efficiency (meaning that the component count of arch. 1 is much less than that of arch. 2), involving only $N_a \ll N_p$ active antennas (at the AMAF) and therefore a much lower number of PAs. 
It is also worthwhile to mention that integrating a large number $N_p$ of patch antennas with their PAs in a single structure, as required by arch. 2 presents non-trivial problems of heat management, which do not exist with arch. 1. All these advantages of the proposed architecture over conventional phased arrays are however highly technology-dependent and therefore more difficult to quantify, and are left here as qualitative comments. 

\section{Conclusions}  
\label{sec:conc}

We have presented a novel RIS-based beamforming architecture with near-field eigenmode feeder. 
The proposed scheme is capable of synthesizing multi-functional beam shapes just by updating the AMAF precoder and the RIS phase-shifters. It has higher power efficiency and significant hardware simplification with respect to a conventional beamforming array of similar performance, due to the fact that it uses only a small number of active antennas. In this work we assumed continuous phase shifters at the RIS. If the RIS uses finite state phase shifters, efficient codebook design for the RIS is an important problem, especially in view of very large RIS arrays. For a fully digital beamforming at the AMAF, the the AMAF precoder can be optimized in conjunction with the SVD of $\Tm$ and the quantized phase shifts at the RIS. Also the investigation of phase-only pattern techniques at the RIS in order to achieve desired BF masks in an interesting research direction left for future work.


\end{document}